\documentstyle[prl,floats,epsbox,twocolumn,aps]{revtex}

\begin{document}
\draft

\twocolumn[\hsize\textwidth\columnwidth\hsize\csname
@twocolumnfalse\endcsname

\title{ \bf Adaptive Sampling Approach to the Negative Sign Problem\\
in the Auxiliary Field Quantum Monte Carlo Method}
\author{Yoshihiro Asai}
\address{Physical Science Division, 
Electrotechnical Laboratory (ETL), \\
Agency of Industrial Science and Technology (AIST), \\
Umezono 1-1-4, Tsukuba, Ibaraki 305, Japan}
\date{Received on 9 Dec 1998}
\maketitle

\widetext

\begin{abstract}
We propose a new sampling method to calculate the ground state of
interacting quantum systems. This method, which we call the adaptive
sampling quantum monte carlo (ASQMC) method utilises information
from the high temperature density matrix derived from the monte
carlo steps.
With the ASQMC method, the negative sign ratio is greatly reduced
and it becomes zero in the limit $\Delta \tau$ goes to zero even 
without imposing any constraint such like the constraint path (CP)
condition. Comparisons with numerical results obtained by using other
methods are made and we find the ASQMC method gives accurate results
over wide regions of physical parameters values.
\end{abstract}

]

\narrowtext


The negative sign problem in quantum monte carlo simulations has
been an extremely serious problem. In the case of the auxiliary field
quantum monte carlo (AFQMC) method ~\cite{Blanken,Scalapino1,Loh1}
applied to the two-dimensional (2D) Hubbard model, we always have this 
problem in the ground state and in low temperature regions except at
half-filling, where we have particle-hole symmetry.~\cite{Hirsch}
The negative sign difficulty is somewhat reduced and calculations become
more feasible only when the filling is far away from half-filling, and/or
when we use a smaller value of $U/t$, and/or when we use a smaller value
of the inverse temperature $\beta$ or the projecting time $\tau$.  
Such a reduction is also observed when the electron filling is such that
the corresponding non-interacting model has a closed-shell electronic
structure and the system size is small. The quantum monte carlo 
results on the 2D models~\cite{Scalapino2,Loh2,White,Sorella,Imada}
and the Hubbard ladder model~\cite{two-chain} obtained so far are in
regimes where the above conditions are satisfied.
However, to get more insight into the physics of strongly correlated
electrons, it is highly desirable to develop more robust numerical
methods whose applicability and accuracy do not depend on the details
of the system to be studied.

Recently, Zhang, Carlson and Gubernatis developed the constrained
path quantum monte carlo (CPQMC) method.~\cite{Zhang}
Their method consists of the two ideas: the random walker in the
configuration space (RWCS) and the constrained path (CP) conditions, 
which is local.  The latter is a variant of the nonlocal positive
projection condition proposed by Fahy and Hamman~\cite{Fahy} but is
simpler to impose. Both of their methods are variational but the rate of
convergence to the ground state of the CPQMC method against a variation of
the trial wavefunction has been reported to be very fast in weak and
intermediate regions of $U/t$ and the closed shell case.
We will demonstrate that the CP condition is not necessary to reduce the
negative sign ratio, if we adopt the adaptive sampling method which we
propose here in the standard projector AFQMC (PAFQMC) algorithm.
 
In applying the PAFQMC method to the Hubbard model: 
$H = T + V_2 $, $T = -t \sum_{\langle i,j \rangle s}
( c^\dagger_{i s} c_{j s} + H.C. )$,  $V_2 = U \sum_{i} 
n _{i \uparrow} n _{i \downarrow}$, where $\langle i,j \rangle$ 
indicates the sum is taken over the pairs of nearest-neighbor sites, 
we use the density matrix: 
$\langle \Psi_t | \exp{ (-\tau H) } | \Psi_t \rangle$.
The density matrix is expressed by using the Suzuki-Trotter (ST) formula
and the Stratonovitch-Hubbard (SH) transformation in the following way:
\begin{eqnarray}
\langle \Psi_t | \exp{ (-\tau H) } | \Psi_t \rangle =
Tr_{\sigma(1), \sigma(2), \cdots , \sigma(L)}
\langle \Psi_{t \uparrow} | B_{\uparrow}(\sigma(L)) \cdots \nonumber \\
B_{\uparrow}(\sigma(1)) |\Psi_{t \uparrow} \rangle
\times
\langle \Psi_{t \downarrow} | B_{\downarrow}(\sigma(L)) \cdots
B_{\downarrow}(\sigma(1)) |\Psi_{t \downarrow} \rangle , \nonumber
\end{eqnarray}
where $L = \tau/\Delta \tau$ and $\Delta \tau$ is the discretized
projecting time $\tau$,
$B_{\alpha} (l) = \exp{ (-\Delta \tau T) }
\exp{ (-\Delta \tau V_{1 \alpha} (l) )}$ 
and
$
V_{1 \alpha} (l)
= \delta_{i j} ( \alpha a_U \sigma_i(l) - (1/2) \Delta \tau U ) .
$
In the $V_{1 \alpha} (l)$ factor $a_U$ is defined to be
$tanh ^{-1} \sqrt { tanh (\Delta \tau U/4) }$ and $\alpha=+1$
for up spins and $-1$ for down spins.
Hereafter, we denote 
$U_{\sigma}(\tau,0) = B(\sigma(L)) \cdots B(\sigma(1)) $ and
the spin indices will be suppressed. Products of both up-spin
and down-spin elements are implicit throughout this letter.
$\tau$ is the total projection time.
The trace over the Ising SH fields:
$\sigma(1), \sigma(2) \cdots, \sigma(L)$
is achieved by using the importance
sampling method. 
The site index $i$ of the SH fields is suppressed here.
After we take the trace over the SH fields, we 
can calculate any ground state expectation value and the ground 
state wavefunction:
$
|\Psi_{exact} \rangle \simeq  
Tr_{\sigma(1), \sigma(2), \cdots , \sigma(L)} 
B(\sigma(L)) \cdots B(\sigma(1)) |\Psi_{t \uparrow} \rangle .
$ 
There remains freedom how we take the trace. In the standard AFQMC
method, the trace over the SH fields
is taken simultaneously and we use the weight function:
$
P = | \langle \Psi_t | U_{\sigma}(\tau , 0)  | \Psi_t \rangle | .
$
There remains another choice in how we take the trace.
If we define:
$
|\Psi_i \rangle =
Tr_{\sigma(i)} B(\sigma(i)) |\Psi_{i-1} \rangle ,
$
and $\Psi_0 = \Psi_t$,
\begin{eqnarray}
|\Psi_{exact} \rangle \simeq
Tr_{\sigma(2), \cdots , \sigma(L)} B(\sigma(L)) \cdots B(\sigma(2))
|\Psi_1 \rangle \nonumber \\ 
= Tr_{\sigma(3), \cdots , \sigma(L)} B(\sigma(L)) \cdots B(\sigma(3))
|\Psi_2 \rangle \nonumber \\
= \cdots 
= Tr_{\sigma(L)} B(\sigma(L)) |\Psi_{L-1} \rangle . \nonumber
\end{eqnarray}
We can then take the trace sequentially. This is similar to the
propagation process in RWCS and both this method and the standard
AFQMC method should give identical results, if proper monte carlo 
sampling is done. 

To take the sequential trace properly, we introduce the adaptive
sampling method and we define the weight function as follows:
\begin{displaymath}
P_{\tau '} =
|
\langle \Psi_t | U_{\sigma}(\tau ,\tau ') | \Psi_t \rangle
\langle \Psi_t | U_{\sigma}(\tau ', 0) | \Psi_t \rangle | ,
(0 \le \tau ' \le \tau) ,
\end{displaymath} 
rather than the standard weight function:
$
P = | \langle \Psi_t | U_{\sigma}(\tau , 0)  | \Psi_t \rangle | .
$
$\tau '$ is an imaginary time whose SH field is tried to be updated
and $\tau '$ changes during the simulation;
the SH field to be updated belongs to the $l= \tau '/ \Delta \tau$-th 
imaginary-time slice.
We use the local-flip update scheme and the field is updated
from $l = 1$ to $l = L$ ( from $\tau ' = \Delta \tau$ to $\tau ' = \tau$).
The ratio used in judging to accept or to reject the update of 
the SH field of the i-th site and of
the $l= \tau '/ \Delta \tau$ -th imaginary-time slice is given as 
follows:
\begin{displaymath}
r_{\tau '} = 
| \langle \Psi_t | U_{-\sigma_i(\tau ')}(\tau ', 0) | \Psi_t \rangle /
  \langle \Psi_t | U_{+\sigma_i(\tau ')}(\tau ', 0) | \Psi_t \rangle | .
\end{displaymath}
We use the heat bath method and the acceptance probability is given by
$R_{\tau '} = r_{\tau '}/ ( 1 + r_{\tau '} )$. 
The $\tau '$ dependent weight function is the unique feature
of our adaptive sampling method.
The update of the SH field is initiated by using the "high
temperature density matrix" (the projecting time 
$\tau' = \Delta \tau$ ). 
All the $N$ SH fields on the imaginary-time $\tau '$ are tried to be 
updated, where $N$ is the number of sites. The series of trial updates
on the imaginary-time $\tau'$ is repeated $M$ times. 
After the trials have completed,
$\tau '$ increases by $\Delta \tau$. The procedure is 
repeated until $\tau'$ becomes $\tau$. All these constitute a sweep. 
The next sweep starts with $\tau ' = \Delta \tau$, again. 
We call our method the adaptive sampling quantum monte carlo (ASQMC)
method.

Measurements may be done with the re-weighting function:
$
Rw(\tau ') = | W_{\sigma} / \tilde{W}_{\sigma}(\tau ') | ,
$
$
W_{\sigma} = 
\langle \Psi_t | U_{\sigma}(\tau , 0) | \Psi_t \rangle ,
$
$
\tilde{W} _{\sigma}(\tau ') =
\langle \Psi_t | U_{\sigma}(\tau ,\tau ') | \Psi_t \rangle
\langle \Psi_t | U_{\sigma}(\tau ', 0) | \Psi_t \rangle .
$
Let $O$ be one-body operators such like: 
$
O = c^{\dagger}_{\vec{k} s} c_{\vec{k} s} ,
$
where $\vec{k}$ is the wave vector and $s$ is the spin variable.
The expectation value of $O$ can be calculated as follows:
\begin{displaymath}
\langle \langle O \rangle \rangle = 
\end{displaymath}
\begin{displaymath}
\frac{Tr_{\sigma} \langle O \rangle Sign W_{\sigma} Rw(\tau ')
(|\tilde{W}_{\sigma}(\tau ')|/Tr_{\sigma} |\tilde{W}_{\sigma}(\tau ')|)}
{Tr_{\sigma} Sign W_{\sigma} Rw(\tau ')
(|\tilde{W}_{\sigma}(\tau ')|/Tr_{\sigma} |\tilde{W}_{\sigma}(\tau ')|)}
\end{displaymath}
We may make measurements only by use of configurations of the SH field
obtained just after the trial updates at $\tau ' = \tau$ ($l = L$) and
then the re-weighting function $Rw(\tau) = 1$. 
In this case, we may evaluate the expectation value as 
follows:
\begin{displaymath}
\langle \langle O \rangle \rangle = 
\frac{\sum_{ns} \langle O \rangle Sign W_{\sigma}}
{\sum_{ns} Sign W_{\sigma}}
\end{displaymath}
where
\begin{displaymath}
\langle O \rangle = 
\frac{ 
\langle \Psi_t | O 
U_{\sigma}(\tau , 0) | \Psi_t \rangle }
{\langle \Psi_t | U_{\sigma}(\tau , 0) | \Psi_t \rangle} ,
\end{displaymath}
and $n_s$ is the number of samples.
For the two-body (or more) physical quantities, we use the Wick 
theorem. 
The measurements in this form are called a mixed estimator:
$
\langle \Psi_t | O | \Psi_{exact} \rangle .
$
One of the difficulties of the ASQMC method is measurements
because the mixed estimator does not give exact expectation
values other than for energy and sign.
To remedy the situation, we may use the following weight function:
\begin{displaymath}
P = | \langle \Psi_t | U_{\sigma}(\tau , 0)  | \Psi_t \rangle |  
(0 \le \tau ' \le \tau_c) ,
\end{displaymath}
\begin{displaymath}
P_{\tau '} =
|
\langle \Psi_t | U_{\sigma}(\tau ,\tau ') | \Psi_t \rangle
\langle \Psi_t | U_{\sigma}(\tau ', 0) | \Psi_t \rangle | ,
(\tau_c \le \tau ' \le \tau) .
\end{displaymath} 
Measurements in this case are made only when
$ 0 \le \tau ' \le \tau_c $ .
In this interval $Rw(\tau ') = 1$ and measurements such as:
\begin{displaymath}
\langle \langle O \rangle \rangle = 
\frac{\sum_{ns} \langle O \rangle Sign W_{\sigma}}
{\sum_{ns} Sign W_{\sigma}}
\end{displaymath}
where
\begin{displaymath}
\langle O \rangle = 
\frac{ 
\langle \Psi_t | U_{\sigma}(\tau, \tau ') O 
U_{\sigma}(\tau ', 0) | \Psi_t \rangle }
{\langle \Psi_t | U_{\sigma}(\tau , 0) | \Psi_t \rangle} .
\end{displaymath}
may be done there.
We call this measurement the standard measurement.

Instead of using $\tilde{W}_{\sigma}(\tau ')$ to define $Rw(\tau ')$,
we may be able to use any weight function which does not necessary
have any physical correspondence to the original problem.
For example, we may use the weight function of the half-filled
Hubbard model $\Omega_{\sigma}$ to study the doped Hubbard model.
The re-weighting function in this case is 
$Rw = |W_{\sigma}/\Omega_{\sigma}|$.
The expectation value of any one-body operator is given:
\begin{displaymath}
\langle \langle O \rangle \rangle = 
\end{displaymath}
\begin{displaymath}
\frac{Tr_{\sigma} \langle O \rangle Sign W_{\sigma} Rw
(|\Omega_{\sigma}|/Tr_{\sigma} |\Omega_{\sigma}|)}
{Tr_{\sigma} Sign W_{\sigma} Rw
(|\Omega_{\sigma}|/Tr_{\sigma} |\Omega_{\sigma}|)}
\end{displaymath}
As the re-weighting procedures do not introduce any approximation
to the theory, both of the mathematical expressions for the expectation
value of operator $O$ are exact.
The numerical effectiveness of the re-weighting methods depends on 
whether $Sign W_{\sigma}$ is mostly positive and if the re-weighting
function $Rw$ is not very small. Unless the two conditions are satisfied,
numerical application of re-weighting methods will not be successful.

\begin{figure}[htbp]
\epsfxsize=9.0cm
\epsffile{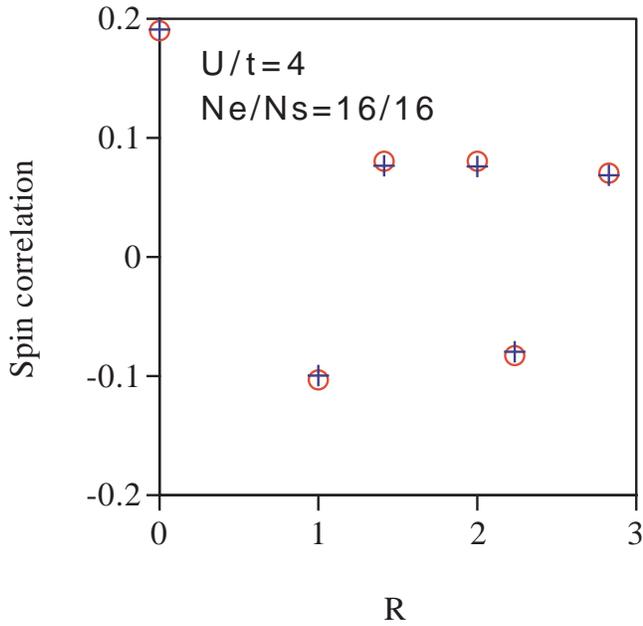}
\vspace*{1.0cm}
\caption{The spin-spin correlation function as a function
of the distance $R$ calculated by the ASQMC and the standard AFQMC 
methods at half-filling.
The calculations were made on the $4 \times 4$ lattice. $U/t=4$.
Open circles and crosses are data obtained by the
ASQMC and the standard AFQMC methods, respectively.}
\label{fig:spin-spin}
\end{figure}

The ergodicity is guaranteed in our method. The ASQMC method is one of the
exact algorithms like the standard AFQMC method and some other
re-weighting methods applied to it, irrespective of the sign problem. This
can be checked numerically at half-filling, where there is no sign problem
and comparisons of results obtained by using the ASQMC method and the
standard AFQMC method are possible. We compared calculated results of the
spin-spin correlation function: 
$
S(\vec{q} ) = 
1/N \sum_{ i,j }\exp{ [-i \vec{q} \cdot (\vec{r}_i -\vec{r}_j )] }
S^z_i S^z_j 
$
of the 4 $\times$ 4 Hubbard model at the half-filling.
The results are Fourier transformed into the real space and they are 
plotted against the distance $R$ in Fig. \ref{fig:spin-spin}. 
We find nice agreement over all the distances within the cluster.

With our adaptive sampling method, we improve the trial wavefunction 
$\Psi_i (i=0,1,\cdots, L-1)$ sequentially so that the final short time
projection:
$
\exp(-\Delta \tau H) |\Psi_{L-1} \rangle
$
extracts out the ground state and therefore ASQMC method is expected 
to be less affected by the negative sign problem. Our method is easily 
implemented by modifying the standard AFQMC code.
We do not use the population control procedure
in our implementation so that $\Delta \tau$ needs to be small.
Here, we have studied the minimal case $M=1$.

\begin{figure}[htbp]
\epsfxsize=8.5cm
\epsffile{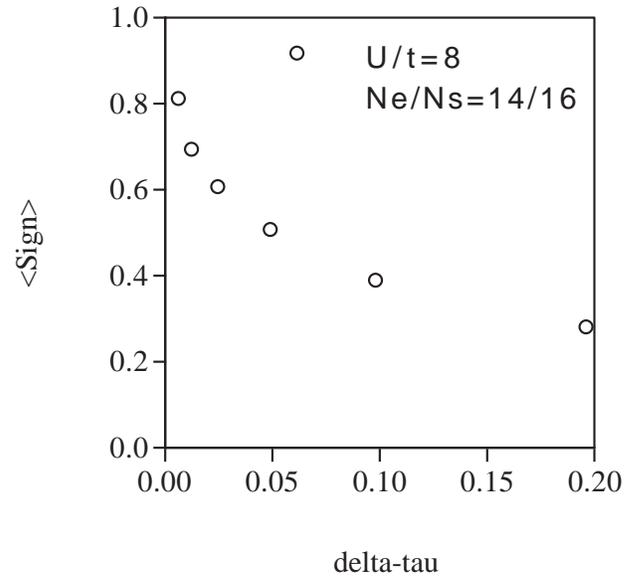}
\vspace*{1.0cm}
\caption{The sign expectation value $\langle Sign \rangle$
as a function of $\Delta \tau$ calculated with the ASQMC method.
We put $\tau_c=0$ and $U/t=8$ and $\rho=0.875$. 
$\tau/t=10$.
The calculation is made
on the $4 \times 4$ Hubbard model.}
\label{fig:signs}
\end{figure}

The benefit of our ASQMC method is that it greatly reduces the negative 
sign ratio without imposing any constraint such like the CP condition.
We calculated the expectation value of the sign 
$\langle Sign \rangle$ as a function of $\Delta \tau$.
$\tau_c$ was set to be zero and we used the mixed estimator method
to calculate the sign expectation value $\langle Sign \rangle$.
We used the projecting time $\tau/t = 10$.
Calculations were made with various physical parameters values
of the 4 $\times$ 4 
Hubbard model, but we show the result obtained with $U/t = 8$
and with the density $\rho = N_e/N_s =0.875$, where there is
very serious negative sign problem when we use the standard
AFQMC method.
The result is shown in Fig. \ref{fig:signs}. 
When $\Delta \tau \to 0$, $\langle Sign \rangle \to 1$.
We have no negative sign problem in the limit of 
$\Delta \tau \to 0$. Over a wide range of finite $\Delta \tau$,
the negative sign ratio obtained with our ASQMC method
is much reduced in comparison with the standard AFQMC method. When we use 
smaller value of $U/t$ or when we study the closed shell filling case, 
$\langle Sign \rangle$ tends to become much closer to $1$ with the
same value of $\Delta \tau$. 
The rate of convergence
to $\langle Sign \rangle \to 1$ depends on values of physical 
parameters.   

Our ASQMC method is expected to give accurate numerical results,
because $Sign W_{\sigma}$ is mostly positive and $Rw(\tau ') = 1$ .
( Measurements are done either at $\tau ' = \tau $ 
or $ 0 \le \tau ' \le \tau_c $ . )
We prefer to use the standard measurement here to reduce the
$ \Delta \tau $ error, which comes not only from
the Trotter error but also from the finite step size $\Delta \tau$
used in taking the sequential trace
of the auxiliary fields.
The ground state wavefunction obtained by using finite $\Delta \tau$
may be expanded by powers of $ \Delta \tau $:
$
\Psi ( \Delta \tau ) \simeq
\Psi_{exact} + 
\Delta \tau \Psi ' (0) ,
$
where
$
\Psi ' (0) =
( \partial / \partial \Delta \tau ) \Psi ( \Delta \tau ) |_{\Delta \tau = 0}
$
and 
$\Psi ' (0)$ is orthogonal to $\Psi_{exact}$.
If we use the mixed estimator measurement, we obtain the expression:
$
\langle \Psi_t | H | \Psi (\Delta \tau) \rangle =
E_{exact} \langle \Psi_t | \Psi_{exact} \rangle +
\Delta \tau \langle \Psi_t | H | \Psi '(0) \rangle .
$
On the other hand if we use the standard measurement, we get :
$
\langle \Psi (\Delta \tau) | H | \Psi (\Delta \tau) \rangle =
E_{exact} \langle \Psi_{exact} | \Psi_{exact} \rangle +
\Delta \tau^2 \langle \Psi ' (0) | H | \Psi ' (0) \rangle ,
$
where $E_{exact}$ is the exact ground state energy of the 
Hamiltonian $H$.
We notice that larger $\Delta \tau$ error remains in the
mixed estimator measurements than in the standard measurements, which
is also observed in numerical simulations.
So the standard measurement is the better choice even when one calculates
energies.
The negative sign ratio obtained by using the standard measurement
method is somewhat larger than that obtained by using the mixed
estimator method, but it is still far more reduced than that obtained
by using the standard AFQMC algorithm.
Hereafter throughout this article, we adopt the standard measurement
method in our ASQMC calculations.

We compared calculated total energies, one and two-body correlation functions
obtained by using the ASQMC method with those obtained by using the
Lanczos exact diagonalization (ED) method and other quantum monte carlo
methods.
Values of $\Delta \tau$ used are $0.025$ and $0.0125$.
A 5000 sweep run on the $8 \times 8$ lattice takes almost 80 hours
of CPU time on the Alpha workstation with the Alpha 21164 / 533MHz
CPU chip.
We found the ASQMC method gives accurate results over wide range
of physical parameters values on $4 \times 4$, $6 \times 6$,
and $8 \times 8$ lattices. In the following paragraph, we show results of
such comparisons.

\begin{figure}[htbp]
\epsfxsize=8.5cm
\epsffile{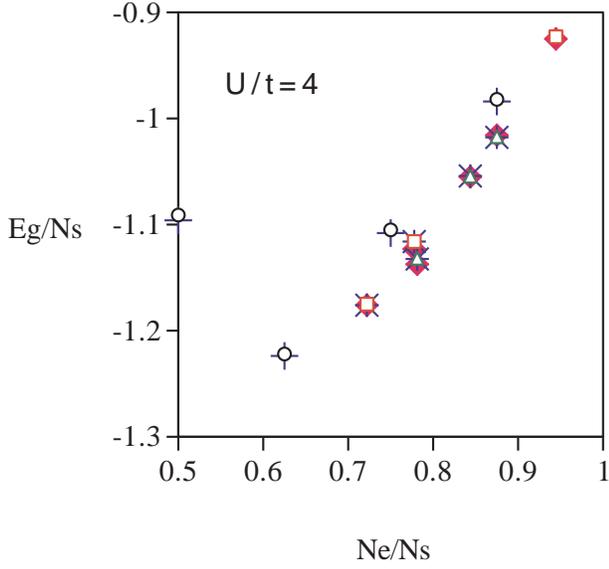}
\vspace*{1.0cm}
\caption{Total energies per site of the two dimensional Hubbard model with
various system size calculated by the ASQMC, ED, the CPQMC, and the AFQMC
methods. $t=1$ and $U=4$.
We plot the energies as a function of electron density $\rho$.
Open circles, open squares, and open triangles denote
results obtained by the ASQMC method with $4 \times 4$, $6 \times 6$,
and $8 \times 8$ lattices, respectively.
Crosses, asterisks, and closed diamonds
denote ED, CPQMC, and AFQMC results, respectively.}
\label{fig:energ1.U=4}
\end{figure}

\begin{figure}[htbp]
\epsfxsize=9.0cm
\epsffile{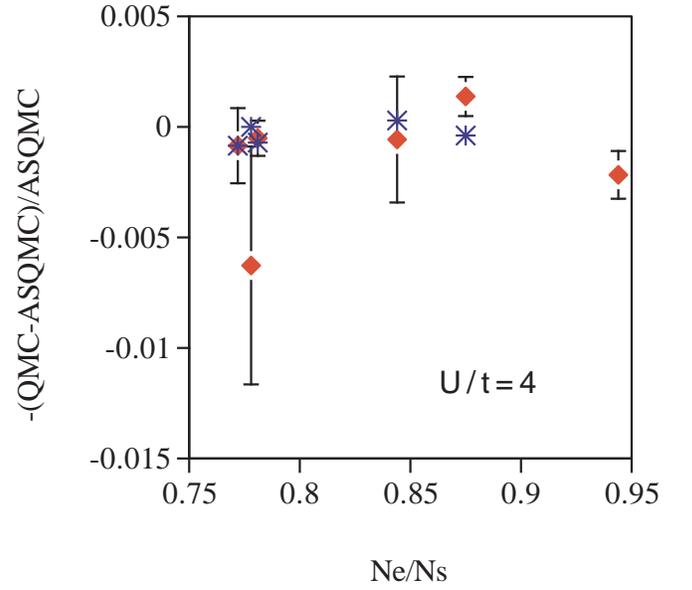}
\vspace*{1.0cm}
\caption{Relative errors of total energies per site calculated by the
CPQMC and the AFQMC methods. They are compared with the ASQMC results.
The physical parameters and fillings are the same as Fig.3. Crosses and
closed diamonds denote the relative errors of the CPQMC and the AFQMC
results, respectively.}
\label{fig:energ2.U=4}
\end{figure}

First, we compare total energies of $U/t=4$ Hubbard model
of various lattice size: $4 \times 4$, $6 \times 6$, $8 \times 8$.
Fillings $\rho$ are such that both the closed shell case and open shell
case are included. 
Agreements of the total energies calculated by using the ASQMC method
with those obtained by using ED,~\cite{ED} 
the CPQMC,~\cite{Zhang} and the AFQMC~\cite{Imada}
methods are rather nice up to 3 digits over the wide range of fillings
and lattice size. This is shown in Fig. \ref{fig:energ1.U=4}.
To see this in more detail, we have plotted relative errors of the
total energies in Fig. \ref{fig:energ2.U=4}. The CPQMC results
agrees with our ASQMC results within $0.1\%$ of errors in the
all cases studied here, but the AFQMC results does not always agree
with our ASQMC results very precisely. The largest error we
found is $0.5\%$. It seems that the extrapolation procedure used to
make inferences of the ground state energies without taking in to account
of statistical errors described in Ref.~\cite{Imada}
does not work so well.

\begin{figure}[htbp]
\epsfxsize=9.0cm
\epsffile{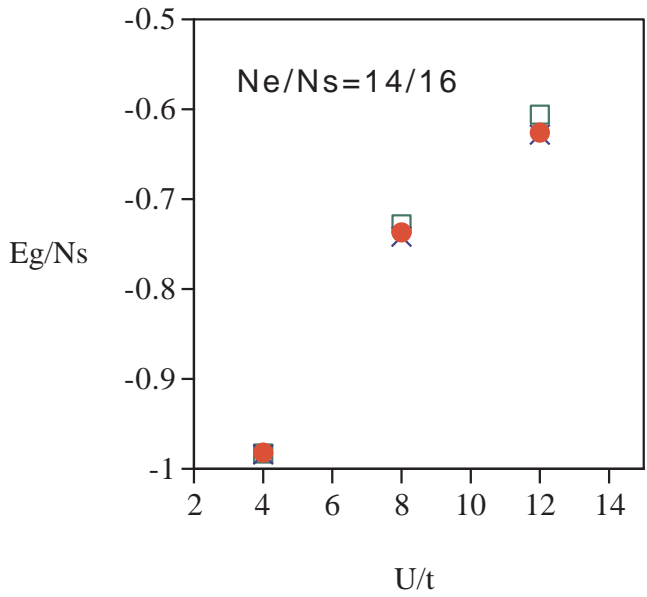}
\vspace*{1.0cm}
\caption{Total energies per site as a function of $U/t$ calculated
by the ASQMC, CPQMC and ED methods. The calculations were
made on the $4 \times 4$ lattice and $\rho = 14/16$.
Closed circles, crosses, and open squares denote
ASQMC, ED, and CPQMC results, respectively.}
\label{fig:energ1.4x4}
\end{figure}

\begin{figure}[htbp]
\epsfxsize=9.0cm
\epsffile{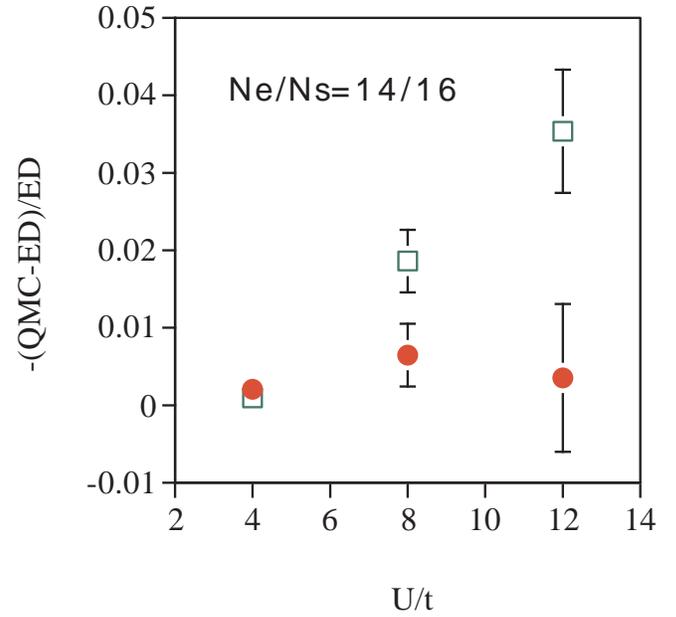}
\vspace*{1.0cm}
\caption{Errors of total energies per site as a function of $U/t$.
The physical parameters and fillings are the same as Fig.5. Closed circles
and open squares are errors of the ASQMC and the CPQMC results,
respectively.}
\label{fig:energ2.4x4}
\end{figure}

We next compare total energies of the $4 \times 4$ Hubbard model
for various $U/t$ values calculated by using the ASQMC, the
CPQMC~\cite{Zhang} and ED methods.~\cite{ED} 
We set the filling $\rho= 14/16$.
Agreements between the ASQMC and ED results are very good
but we found small systematic increase of deviation of the
CPQMC results from the other two results, as we increase $U/t$.
To see in more detail this tendency found in Fig. \ref{fig:energ1.4x4},
we have plotted relative errors of the total energies in 
Fig. \ref{fig:energ2.4x4}. While the CPQMC results deviate from the
ED results systematically, the ASQMC results do not.
Because we do not impose any constraint such like the CP condition,
the ASQMC method does not have any systematic bias even in the
large $U/t$ region.
This demonstrates that our ASQMC method is superior to the CPQMC method
in the large $U/t$ region.

\begin{figure}[htbp]
\epsfxsize=9.0cm
\epsffile{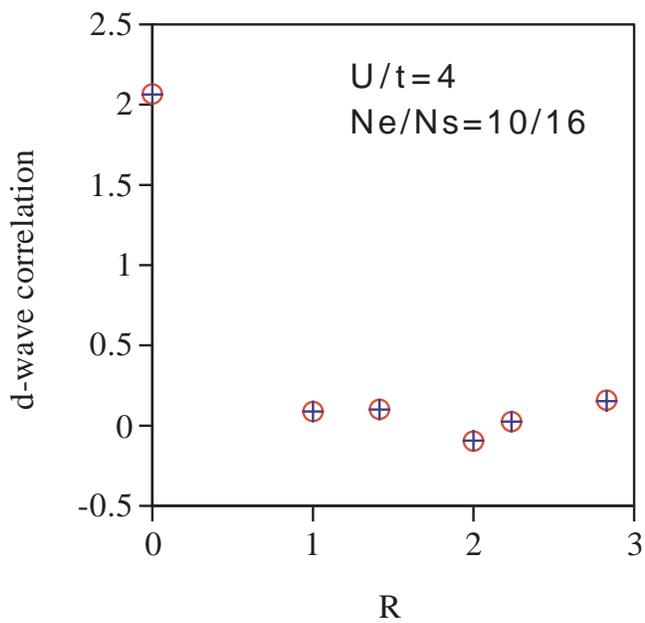}
\vspace*{1.0cm}
\caption{The d-wave superconducting correlation function as a function
of the distance $R$ calculated by the ASQMC and ED methods.
The calculations were made on the $4 \times 4$ lattice. $U/t=4$ and
$\rho = 10/16$. Open circles and crosses are data obtained by the
ASQMC and ED methods, respectively.}
\label{fig:d-wave}
\end{figure}

To provide other benchmark of our ASQMC method, we have calculated
the d-wave pairing correlation function of the 4 $\times$ 4 Hubbard
model defined by:
$
<O(R)O^{\dagger}(0)>
$
with
$
O^{\dagger}(R) = 
    c^{\dagger}_{\uparrow} (R) c^{\dagger}_{\downarrow} (R+x)
  + c^{\dagger}_{\uparrow} (R) c^{\dagger}_{\downarrow} (R-x)  
  - c^{\dagger}_{\uparrow} (R) c^{\dagger}_{\downarrow} (R+y)
  - c^{\dagger}_{\uparrow} (R) c^{\dagger}_{\downarrow} (R-y)  
$
and compared it with the result obtained with the Lanczos diagonalization
method.~\cite{Moreo}
The result is shown in Fig. \ref{fig:d-wave}.
We again, obtained a good agreement with the Lanczos diagonalization 
result over all the distances within the cluster studied.

As far as the present author knows, there are no other reliable
numerical methods than ED, the AFQMC and the CPQMC methods
to compare with our ASQMC methods for the 2D Hubbard model.
Our ASQMC results always agree well with the best of the results
obtained by using the other methods.
Thus our ASQMC method turns out to be a very accurate method.
The ASQMC method will be useful to study a wider physical parameter 
region including the large $U/t$ region, 
many more cases of filling and band structures, including the open
shell cases, and larger
lattice sizes that have not yet been explored by the quantum monte
carlo methods.

To summarize, we have proposed the adaptive sampling method to utilize
information from the "high temperature density matrix" in the
thermalization process of monte carlo steps to calculate the ground state.
With the adaptive sampling method, the negative sign ratio decreases 
to 0 when $\Delta \tau \to 0$ without imposing any constraint such as the
CP condition.
Over a wide range of finite $\Delta \tau$, the negative
sign ratio is far more reduced than that in the standard AFQMC method.
We compared calculated energies and two-body correlation functions 
obtained by using our method and with those obtained by using the 
Lanczos diagonalization method and other quantum monte carlo methods
found in the literature and we found the ASQMC method gives accurate
results over a wide range of physical parameters values.

The author would like to appreciate to Dr. J.E. Gubernatis and 
Prof. J.R. Schrieffer for conversations made during the author's
visit to their institutes on the subject discussed in this letter. 
He appreciates Prof. B.A. Friedman for reading this manuscript
and giving some comments.
The work is financially supported by the project E-TK970005 in ETL.
Part of the calculations were made by using facilities of the Research
Information Processing System (RIPS) center of AIST, and the 
Supercomputer Center, Institute for Solid State Physics (ISSP),
University of Tokyo, to which the author would like to thank.


%
%

%
%

\end{document}